\documentclass[a4paper, 12pt]{article}

\pdfoutput=1

\usepackage[english]{babel}
\usepackage{cite}
\usepackage{a4wide}
\usepackage{amsmath}
\usepackage[charter,cal=cmcal,greekuppercase=italicized]{mathdesign}
\usepackage{dsfont}
\usepackage[T1]{fontenc}
\usepackage{slashed}
\usepackage{setspace}
\usepackage{graphicx}
\usepackage{xcolor}
\usepackage{soul}
\usepackage[final,          % override "draft" which means "do nothing"
            colorlinks,     % rather than outlining them in boxes
            linkcolor=purple, % override truly awful colour choices
            citecolor=teal, %   (ditto)
            urlcolor=blue  %   (ditto)
            ]{hyperref}

\makeatletter
\DeclareRobustCommand*{\bfseries}{%
  \not@math@alphabet\bfseries\mathbf
  \fontseries\bfdefault\selectfont
  \boldmath
}
\makeatother

\newcommand{\im}{\ensuremath{\mathrm{i}\,}}

\newcommand{\SU}{\ensuremath{\mathrm{SU}}}
\newcommand{\U}{\ensuremath{\mathrm{U}}}

\newcommand{\MeV}{\ensuremath{\mathrm{MeV}}}
\newcommand{\GeV}{\ensuremath{\mathrm{GeV}}}
\newcommand{\TeV}{\ensuremath{\mathrm{TeV}}}

\newcommand{\hc}{\ensuremath{\text{h.\,c.\ }}}

\newcommand{\vev}{\emph{vev}}
\newcommand{\ie}{i.\,e.~}

\newcommand{\eg}{e.\,g.~}

\usepackage[font=small]{caption}
\usepackage{graphicx}

\newcommand{\mue}{\ensuremath{\mu_\text{eff}}}

\begin{document}

\onehalfspacing

\begin{titlepage}

\vspace*{-15mm}
\begin{flushright}
DESY 18-005
\end{flushright}
\vspace*{0.7cm}

\begin{center} {
\bfseries\LARGE
Higgs Inflation at the LC
}\\[8mm]
W.~G.~Hollik
\footnote{\texttt{w.hollik@desy.de}}
\end{center}
\vspace*{0.50cm}
\centerline{\itshape
Deutsches Elektronen-Synchrotron (DESY),}
\centerline{\itshape
Notkestra\ss{}e 85, D-22607 Hamburg, Germany.}
\vspace*{1.20cm}

\begin{abstract}
\noindent
Most cosmological models of inflation are far away from providing a
smoking gun at low energies. A model of Higgs inflation in the
Next-to-Minimal Supersymmetric Standard Model, however, changes the
NMSSM phenomenology drastically and may be well distinguished from the
pure NMSSM or MSSM at a future Linear Collider.  We point out certain
differences of the inflationary model to the ordinary NMSSM and discuss
the Higgs and neutralino/chargino sector in particular to identify the
smoking gun of inflation at electroweak energies.
\end{abstract}

\noindent
\small Talk presented at the International Workshop on Future Linear
Colliders (LCWS2017), Strasbourg, France, 23-27 October
2017. \texttt{C17-10-23.2}

\end{titlepage}

\setcounter{footnote}{0}

\section{Introduction}\label{sec:intro}
Cosmological models that can be tested in the laboratory are typically
very rare. A smoking gun at low energies of a model acting at a very
large scale like the Planck scale requires a precision machine like a
Linear Collider. Precise electroweak observations may then distinguish
between an ordinary extension of the Standard Model (SM) or an extension
which simultaneously grasps cosmological problems. Early universe
inflation is to be seen as a cosmological fact which has to be
addressed. If it is addressed in a way that interrelates the
Planck-scale physics with Fermi-scale physics, such a model will most
probably modify the Higgs sector of the SM. On an economic basis,
employing the SM Higgs field as the inflaton field of cosmology, such a
model can be called minimal. While Higgs inflation in the SM tends to
become ``unnatural'' towards the high scales, see
Ref.~\cite{Einhorn:2009bh}, a more viable candidate is the scale-free
Supersymmetric Standard Model. The scale-free model requires compared to
the Minimal Supersymmetric Standard Model (MSSM) an additional singlet
superfield and thus is known as the Next-to-Minimal Supersymmetric
Standard Model (NMSSM).

The Higgs sector of the NMSSM is characterised by the
\(\mathbb{Z}_3\)-invariant superpotential
\begin{equation}\label{eq:NMSSM}
\mathcal{W}_\text{Higgs} = \lambda \; \hat S\, \hat H_u \cdot \hat H_d +
\frac{\kappa}{3} \; {\hat S}^3,
\end{equation}
where \(\hat H_u\) and \(\hat H_d\) are the \(\SU(2)_L\) doublet Higgs
superfields and \(\hat S\) the singlet superfield. The scalar components
of the doublet fields decompose as
\begin{equation}
H_u = \begin{pmatrix} H_u^+ \\ H_u^0 \end{pmatrix}, \qquad
H_d = \begin{pmatrix} H_d^0 \\ H_d^- \end{pmatrix},
\end{equation}
such that \(H_u \cdot H_d = H_u^+ H_d^- - H_u^0 H_d^0\). This
superpotential has an accidental \(\mathbb{Z}_3\)-invariance under the
transformation
\[\hat\Phi \to e^{i\frac{2\pi}{3} k} \hat \Phi, \quad k \in
\mathbb{Z}, \quad \text{for } \hat\Phi = \hat H_u, \hat H_d, \hat S,\]
such that only trilinear terms are allowed. Especially the \(\mu\)-term
of the MSSM, \(\mu \; H_u \cdot H_d\), is forbidden if the
\(\mathbb{Z}_3\) symmetry is imposed. After electroweak symmetry
breaking, however, the singlet scalar acquires a vacuum expectation value
(\vev) and dynamically induces a \(\mu\)-term via the \(\lambda\)
coupling, which plays the role of an effective higgsino mass term:
\begin{equation} \label{eq:mueff}
\mue = \lambda \langle S \rangle.
\end{equation}

\paragraph{Non-minimal coupling in Canonical Superconformal
  Supergravity}
The implementation of Higgs inflation in superconformal theories follows
the non-minimal coupling of the Higgs field content to supergravity, as
suggested by Ref.~\cite{Einhorn:2009bh}, and comes with a single
dimensionless and holomorphic coupling \(X(\hat\Phi)\):
\begin{equation} \label{eq:cssnonmin}
\mathcal{L} = -6 \int \operatorname{d}^2 \theta \mathcal{E}
\left[ R + X(\hat\Phi) R - \frac{1}{4} \left( \bar{\mathcal{D}}^2 - 8 R
\right) {\hat\Phi}^\dag \hat\Phi + \mathcal{W}(\hat\Phi)
\right] + \hc + \ldots,
\end{equation}
where \(\mathcal{E}\) is the vierbein multiplet, \(R\) the curvature
multiplet and \(\bar{\mathcal{D}}\) a covariant derivative. The chiral
superfields \(\hat \Phi\) shall be any of the fields \(H_u\), \(H_d\) or
\(S\). A realization of such a non-minimal coupling involving the
doublet Higgs fields only can be found to be
\begin{equation} \label{eq:nonmincoup}
X = \chi \; \hat H_u \cdot \hat H_d,
\end{equation}
with a numerical factor \(\chi\). Note that this term breaks the
\(\mathbb{Z}_3\) symmetry of the NMSSM and the superconformal symmetry.

The addition of the superconformal symmetry breaking term changes the
frame function in Jordan frame supergravity and affects the K\"ahler
potential in such a way that the NMSSM superpotential gets modified
\cite{Ferrara:2010yw, Ferrara:2010in}. In Planck units (\(M_P = 1\)),
the frame function \(\Omega = \hat \Phi_i^* \hat\Phi_i - 3\) gets
extended by the \(\chi\)-term to
\begin{equation}
\Omega_\chi = \Omega - \frac{3}{2} \left( X(\hat\Phi) + \hc \right),
\end{equation}
and similarly the K\"ahler potential
\begin{equation}
\mathcal{K} = -3 \log ( -\Omega /3 ) \to
\mathcal{K}_\chi = -3 \log (-\Omega_\chi / 3).
\end{equation}
In the canonical superconformal supergravity (CSS) model, the frame
function is explicitly given by \cite{Lee:2010hj, Ferrara:2010in}
\begin{equation} \label{eq:framefunc}
\Omega_\text{CSS} = -3 + |\hat S|^2 + |\hat H_u|^2 + |\hat H_d|^2
+ \frac{3}{2} \chi \left( \hat H_u \cdot \hat H_d + \hc \right).
\end{equation}
In order to have successful inflation in the NMSSM, however, a
stabilisator term \(\zeta (\hat S \hat{\bar S})^2\) has to be added
\cite{Lee:2010hj, Ferrara:2010in}, which disappears from the low-energy
phenomenology (Planck-suppressed).

The \(\chi\)-term breaks a continuous \(\mathcal{R}\) symmetry and its
discrete \(\mathbb{Z}_3\) subgroup at dimension six \(\sim \chi
\tfrac{\lambda^2 h^6}{M_P^2}\). Much below the Planck scale, the
additional term induces a correction in the superpotential
\begin{equation}
\mathcal{W}_\text{eff} \to \mathcal{W} e^{\frac{3}{2} \chi H_u \cdot
  H_d / M_P^2} \approx \mathcal{W} + \frac{\langle
  \mathcal{W}_\text{hid}\rangle}{M_P^2} \frac{3}{2} \chi H_u \cdot H_d
\equiv \mathcal{W} + \frac{3}{2} \chi m_{3/2} H_u \cdot H_d,
\end{equation}
where the \vev{} of the hidden sector superpotential can be related to
the gravitino mass scale
\begin{equation}
m_{3/2} \approx \frac{\langle \mathcal{W}_\text{hid} \rangle}{M_P^2}.
\end{equation}
Effectively, the superpotential of the NMSSM gets modified by an
additional \(\mu\)-like term,
\begin{equation} \label{eq:muNMSSM}
\mathcal{W}_\text{iNMSSM} = \lambda \; S H_u \cdot H_d +
\frac{\kappa}{3} \; S^3 + \mu \; H_u \cdot H_d,
\end{equation}
with \(\mu = \frac{3}{2} \chi m_{3/2}\). Thus, the effective
higgsino mass term of the NMSSM Eq.~\eqref{eq:mueff} gets shifted by the
contribution from the non-minimal coupling to supergravity leading to
inflation as
\begin{equation} \label{eq:mumueff}
\mue' = \lambda \; \langle S \rangle + \frac{3}{2} \chi m_{3/2}
= \mue + \mu.
\end{equation}
The low-energy smoking gun of Higgs inflation in the superconformal
setup appears to be the NMSSM extended with an MSSM-like \(\mu\)-term
and can be quite well distinguished from either the pure MSSM or NMSSM
as will be discussed in the following. We refer to this model setup as
the inflationary NMSSM, or short iNMSSM.

\section{A short introduction to the iNMSSM}
We consider the NMSSM extended with the additional \(\mu\)-term as
described above only. Its presence can be motivated from a non-minimal
coupling to supergravity and a proceeding transformation in the K\"ahler
potential in such a way that only the term \(\mu\; H_u \cdot H_d\) is
present in the superpotential with \(\mu = \frac{3}{2} \chi
m_{3/2}\). Cosmological observations require the size of this
non-minimal coupling to be
\begin{equation}
\chi \simeq 10^5 \lambda.
\end{equation}
The size of the \(\mu\)-term is then mainly given by the gravitino mass
\(m_{3/2}\) and the \(\lambda\) coupling, which we will assume to be
\(\mathcal{O}(0.1)\) in order to have sizeable NMSSM
effects. Generically, we also assume \(\mu \sim \mathcal{O}(1\,\TeV)\),
which in combination requires rather light gravitinos of \(m_{3/2} \sim
10 \,\MeV\). This might cause the cosmological gravitino problem, see
Ref.~\cite{Moroi:1993mb}. Overabundance of gravitino dark matter,
however, can be constrained by constraining the reheating temperature
after inflation~\cite{Ellis:1984eq}. We assume that the details of the
inflationary model can accommodate this problem as outlined
in~\cite{Ferrara:2010in}.

\paragraph{Soft SUSY breaking in the iNMSSM and the Higgs potential}
The additional \(\mathds{Z}_3\)-breaking \(\mu\)-term may generate an
additional soft Supersymmetry (SUSY) breaking bilinear parameter in a
similar manner as the Higgs \(B_\mu\)-term exists in the MSSM. All
further \(\mathds{Z}_3\)-breaking soft SUSY breaking terms that are in
general allowed, see~\cite{Ellwanger:2009dp} and \cite{Ross:2011xv}, are
assumed to be suppressed~\cite{Lee:2011dya} and cannot be generated to
sizeable amount by radiative corrections. Therefore, the soft SUSY
breaking potential can be summarised as the usual trilinear terms of the
NMSSM \(\sim A_\lambda, A_\kappa\) plus the bilinear term from the
non-minimal coupling to supergravity:
\begin{equation}
V_\text{soft} = \lambda A_\lambda\, S H_u \cdot H_d
+ \frac{1}{3} \kappa A_\kappa\, S^3
+ \frac{3}{2} B_\mu \chi m_{3/2} \left(H_u \cdot H_d + \hc \right).
\end{equation}
The scalar potential for the two doublet and one singlet Higgs fields is
defined according to the rules of SUSY and consists of the \(F\)- and
\(D\)-terms as well as the contribution from SUSY breaking. In
comparison to the \(\mathds{Z}_3\)-symmetric NMSSM, we have the
additional \(\mu\)-term appearing as mass term for the doublet fields
and the bilinear soft breaking term \(\sim H_u \cdot H_d\). The full
Higgs potential is thus given by
\begin{equation}\label{eq:HiPot}
\begin{aligned}
V_\text{Higgs} =& \left[ m_{H_d}^2 + (\mu + \lambda S)^2 \right] |H_d|^2
+ \left[ m_{H_u}^2 + (\mu + \lambda S)^2 \right] |H_u|^2
+ m_S^2 \, |S|^2 \\
&+ \frac{2}{3} \kappa A_\kappa \, S^3
+ \left[ \kappa \, S^2 + \lambda\, H_u \cdot H_d \right]^2
+ 2 \left( B_\mu \mu + \lambda A_\lambda \, S \right) H_u \cdot H_d \\
&+ \frac{g_1^2 + g_2^2}{8} \left( |H_d|^2 - |H_d|^2 \right)^2
+ \frac{g_2^2}{2} |H_d^\dag H_u|^2,
\end{aligned}
\end{equation}
where we assume all parameters to be real.

The potential of Eq.~\eqref{eq:HiPot} can easily provide the observed
phenomenology of electroweak symmetry breaking meaning Higgs \vev{}s of
the neutral doublet Higgs field components, \(\langle H^0_u \rangle =
v_u\) and \(\langle H^0_d \rangle = v_d\), with \(v_u^2 + v_d^2 = v^2 =
(174\,\GeV)^2\) and \(\tan\beta = v_u / v_d\) a free parameter;
additionally, the singlet \vev{} generates the effective \(\mu\)-term of
the NMSSM, \(\mue = \lambda \langle S \rangle = \lambda v_s\). In order
to do so, the soft SUSY breaking masses \(m_{H_d}^2\), \(m_{H_u}^2\) and
\(m_S^2\) are adjusted in such a way, that the minimisation conditions
\begin{equation}
\frac{\partial V_\text{Higgs}}{\partial H^0_d} \bigg|_\text{vev}
=\; 2 m_{H_d}^2 v_d + \ldots, ~
\frac{\partial V_\text{Higgs}}{\partial H^0_u} \bigg|_\text{vev}
=\; 2 m_{H_u}^2 v_u + \ldots, ~
\frac{\partial V_\text{Higgs}}{\partial S} \bigg|_\text{vev}
=\; 2 m_{H_d}^2 v_s + \ldots
\end{equation}
are fulfiled. Solving for the mass parameters is trivial and we obtain
\begin{subequations}
\begin{align}
m_{H_d}^2 &= - (\mu + \mue)^2 - v^2 \lambda^2 \sin^2\beta
- \frac{1}{2} M_Z^2 \cos(2\beta) + a_1 \tan\beta, \\
m_{H_u}^2 &= - (\mu + \mue)^2 - v^2 \lambda^2 \cos^2\beta
+ \frac{1}{2} M_Z^2 \cos(2\beta) + a_1 \cot\beta, \\
m_S^2 &= a_4 - a_5 - a_7 - v^2 \lambda^2 - 2 \mue^2 \left(
  \frac{\kappa}{\lambda} \right)^2,
\end{align}
\end{subequations}
where \(M_Z^2 = \tfrac{g_1^2+g_2^2}{2} v^2\) and the abbreviations
\(a_i\) are defined in Appendix~\ref{app:defin}.

The electroweak breaking conditions have to be taken with great care,
since the potential possesses multiple minima and even if \emph{one
  particular} minimum is selected to be the electroweak (\emph{desired})
vacuum by the above definition, there might be other minima deeper than
the desired vacuum and thus the \emph{true} vacuum, \ie the global
minimum, is not the desired one anymore. At tree-level, the minimisation
can only be done numerically; at the loop-level, the situation even gets
worse and one has to guess suitable starting values for the numerical
routines, which have the potential to miss several of the minima. We
take those constraints at the tree-level seriously and therefore exclude
parameter points leading to a non-standard true vacuum of the
theory. Typically, this global minimum appears at larger \vev{}s for the
fields and thus gets more easily selected by the cosmological history of
the universe after inflation~\cite{Strumia:1996pr}. Since we start with
\vev{}s shortly below the Planck scale after inflation ends, the
universe while cooling down may get stuck in the higher scale vacuum. If
it is a local minimum, one should consider the tunneling to the desired
one. Typically, however, the larger \vev{} vacuum appears to be deeper
than the desired vacuum.

Besides the fact that there are multiple vacua implying alternative
\vev{}s in the Higgs potential, the Higgs mass matrices (defined in
Appendix~\ref{app:defin}) show tachyonic states at the tree-level
depending on the input parameters. Tachyonic states have negative masses
squared and simply invalidate the electroweak expansion point because
the potential at that point appears to be a local maximum (the
``minimisation'' conditions are rather conditions for stationary points
and may also result in maxima or saddle points) and thus pointing
towards the deeper minimum in the tachyonic direction. Actually,
radiative corrections may lift up the potential in some cases leading to
rather light instead of tachyonic states. We take these constraints
nevertheless seriously and exclude tachyonic parameter configurations
irrespective whether radiative corrections lift the masses up or not.

The tachyonic constraints already confine clear portions of parameter
space that remain valid. In addition, vacuum stability considerations
exclude additional parts at the borderline.

\section{Electroweak phenomenology of the iNMSSM}
The phenomenology of the iNMSSM at the electroweak scale deviates
significantly from the usual NMSSM. On the one hand, the number of
states remain the same which may look like the same phenomenology. On
the other hand, the dependence on certain parameters appears to be very
different and the additional \(\mu\)-term changes the interpretation of
the higgsino mass parameter as well as the functional dependence of the
Higgs masses on it.

First of all, the tachyonic selection rule excludes large as well as
very small (\(\lesssim \sqrt{2} v\)) values of \(\mue\). Moreover, both
values \(\mu\) and \(\mue\) appear to be correlated. This can be seen
from Figure~\ref{fig:tachyexcl}. The tachyonic boundaries can be easily
understood from a look at the mass matrices, see
Appendix~\ref{app:defin}, where the small \(\mue\) value sets
\(A_\lambda\) to be large, which sits on the off-diagonal elements and
thus is responsible for a large mixing which potentially drives one
state negative. Similarly, if the combination \(\mu + \mue\) appears to
be large; therefore same signs of \(\mu\) and \(\mue\) are excluded in
most cases. The trilinear soft SUSY breaking parameter \(A_\kappa\)
mainly influences the pseudoscalar singlet-like state. If this one
appears to be tachyonic for small \(A_\kappa\), larger values of this
parameter have the ability to lift this mass up and open up parameter
space that is excluded with a vanishing \(A_\kappa\). This can be
clearly seen in the comparison of the allowed and excluded parameter
space in the \(\mu\)-\(\mue\)-plane shown in Figure~\ref{fig:tachyexcl}.

\begin{figure}
\begin{minipage}{0.5\textwidth}
\includegraphics[width=\textwidth]{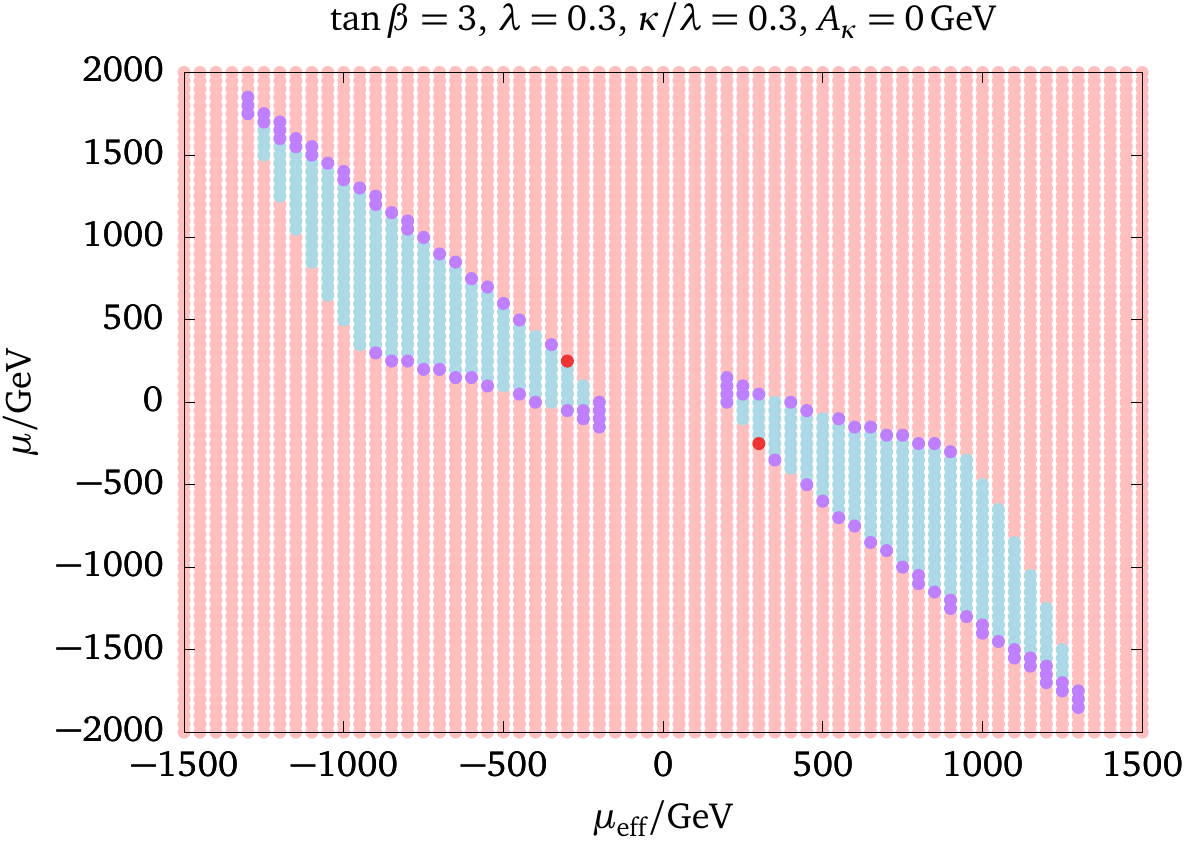}
\end{minipage}%
\begin{minipage}{0.5\textwidth}
\includegraphics[width=\textwidth]{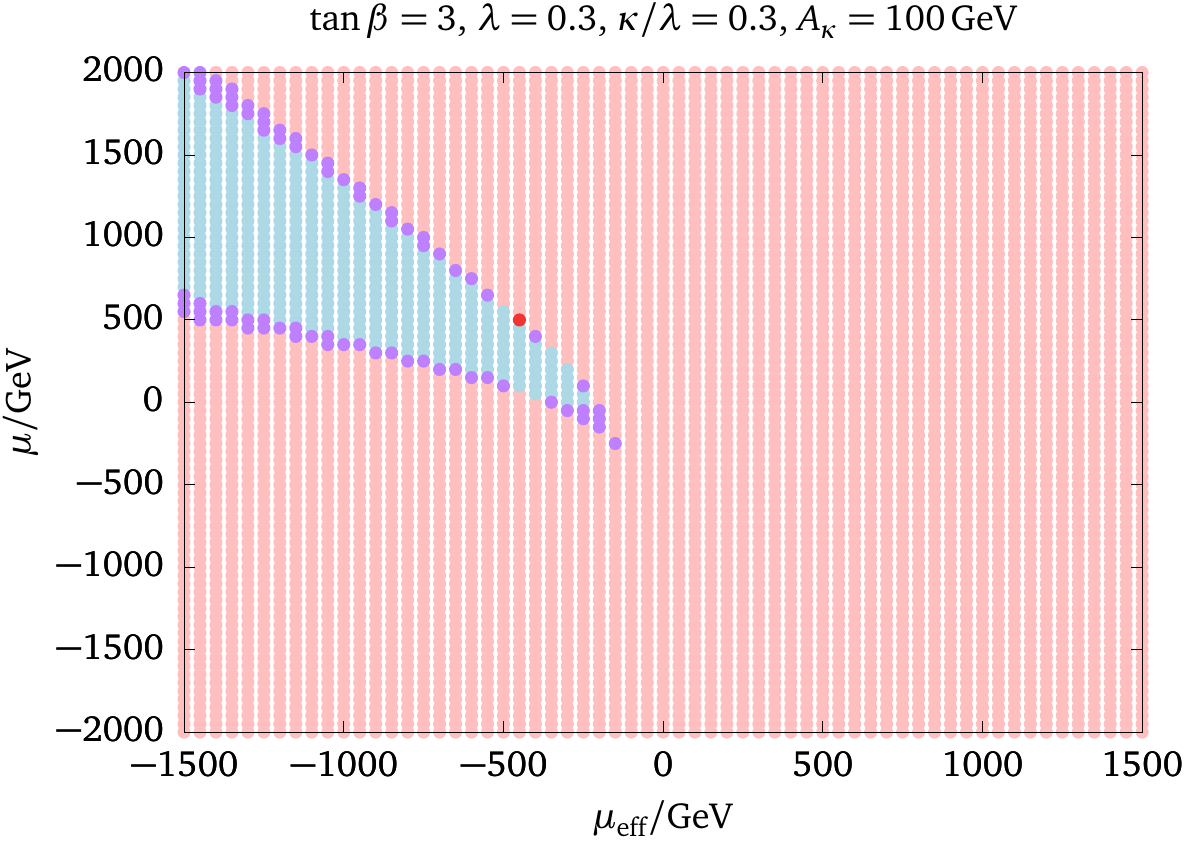}
\end{minipage}
\caption{Mainly tachyonic constraints (pink points) confine the allowed
  parameter space (light blue). In addition, there exist other global
  minima than the electroweak \vev{}s with a mostly long-lived desired
  vacuum (purple) and rarely short-lived configurations (red). On the
  left panel, \(A_\kappa\) is taken to be \(0\,\GeV\), where on the
  right panel \(A_\kappa = 100\,\GeV\). The positive value of
  \(A_\kappa\) disallows the right wing for positive \(\mue\) that was
  allowed for vanishing \(A_\kappa\) (in this case, there is a
  reflection symmetry). Moreover, there is a clear correlation between
  the allowed signs of \(\mu\) and \(\mue\), which in most cases have to
  differ unless \(\mu\) appears to be small. The other sign of
  \(A_\kappa\) reversed the situation.}
\label{fig:tachyexcl}
\end{figure}

\begin{figure}
\begin{minipage}{0.5\textwidth}
\includegraphics[width=\textwidth]{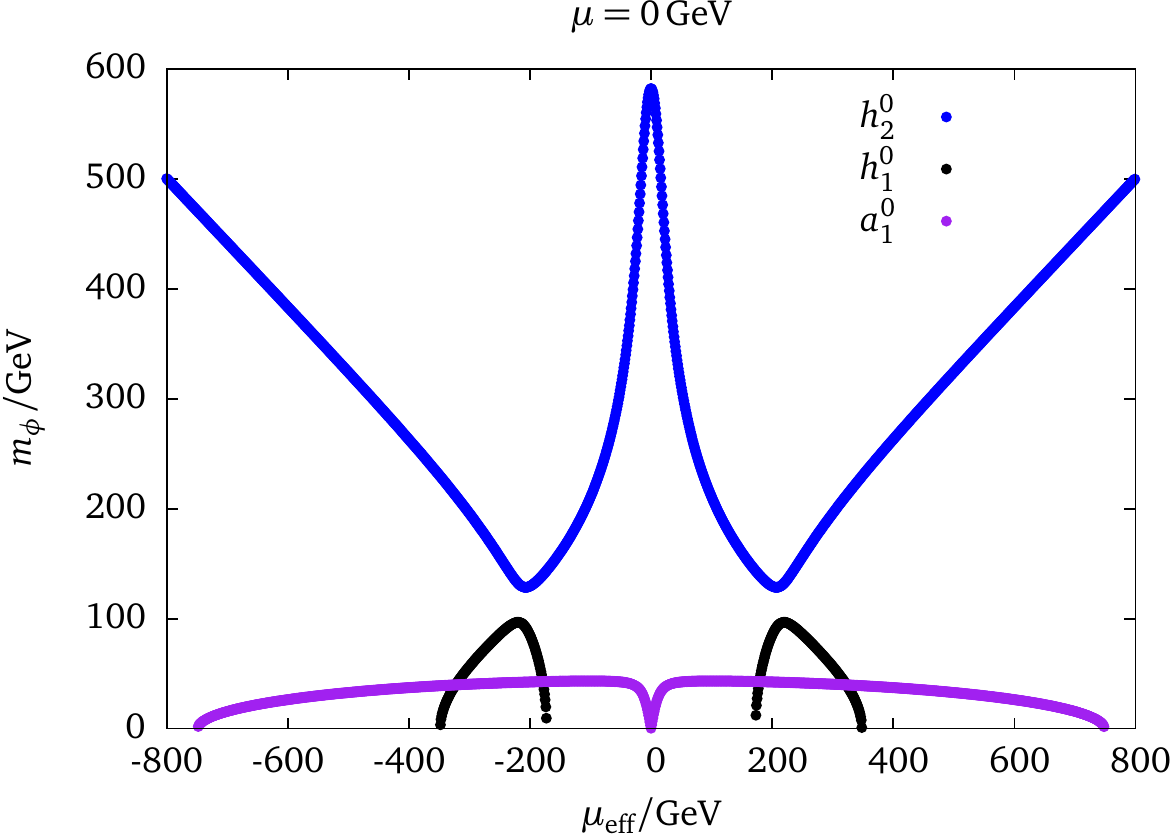}
\end{minipage}%
\begin{minipage}{0.5\textwidth}
\includegraphics[width=\textwidth]{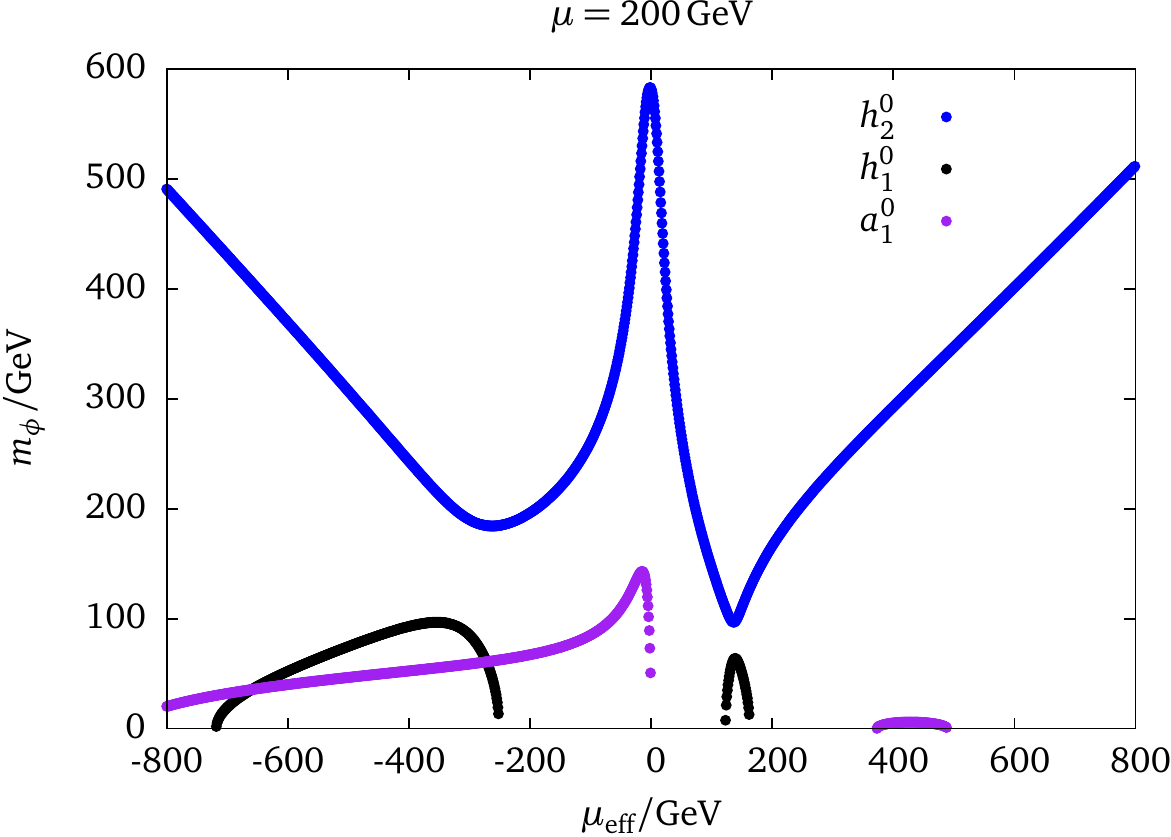}
\end{minipage} \\
\begin{minipage}{0.5\textwidth}
\includegraphics[width=\textwidth]{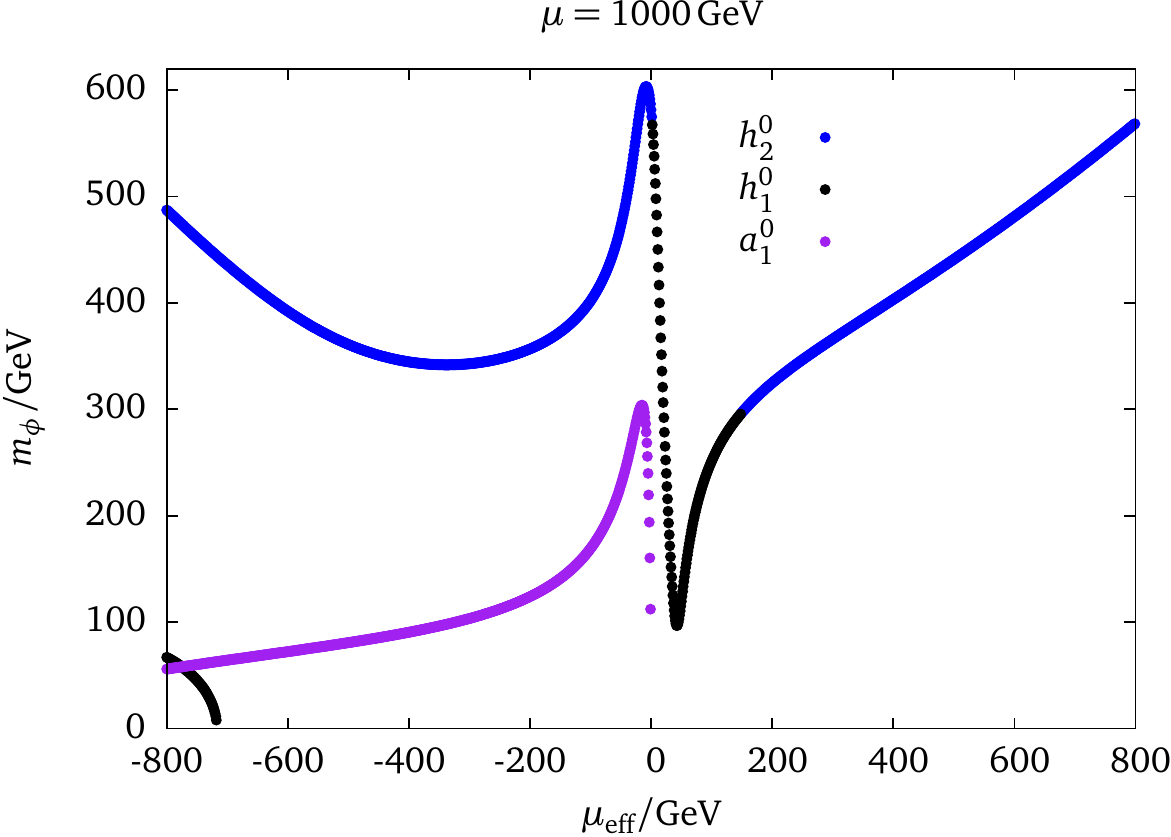}
\end{minipage}%
\begin{minipage}{0.5\textwidth}
\includegraphics[width=\textwidth]{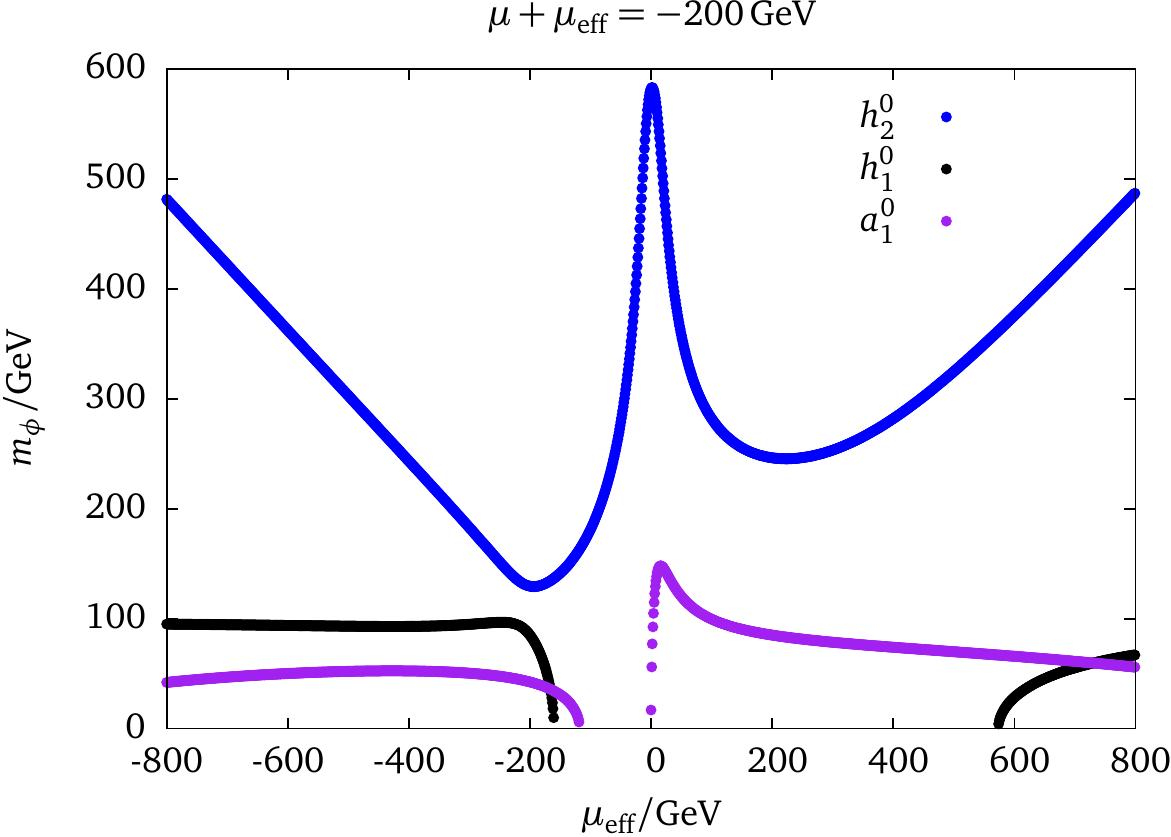}
\end{minipage}
\caption{The spectra of the lightest states and how they vary with
  varying \(\mue\), exemplarily for some choice of parameters. In each
  plot, the \(\mu\) value is fixed to a given value, where the lower
  right plot has a fixed sum \(\mu + \mue = -200\,\GeV\). It can be
  clearly seen which intervals are allowed (those with all three states
  appearing in the plot; where one or more are missing, these are
  tachyonic). In the case with \(\mu = 1000\,\GeV\) \(h_1^0\) and
  \(h_2^0\) apparently change their role which is due to the fact that
  the absolute value of the tachyonic state grows above the
  corresponding value of \(h_2^0\). The scenario with \(\mu =
  200\,\GeV\) shows the feature that the tachyonic exclusions are
  exclusive in the sense that one tachyonic state (scalar or
  pseudoscalar) is enough to exclude the spectrum. Here, both the
  lightest scalar and pseudoscalar have some small interval for positive
  \(\mue\) where they are non-tachyonic but the respective other one is
  and thus all the range for positive \(\mue\) is excluded (where
  \(a_1^0\) turns tachyonic the first time for growing \(\mue\)) and
  additionally already the light scalar mass gets tachyonic at larger
  negative values of \(\mue\). This artefact can be also seen in the
  region plot of Figure~\ref{fig:tachyexcl}.}
\label{fig:Higgsmasses}
\end{figure}

The effect of both \(\mu\) and \(\mue\) on the tachyonicity of states
can be seen from the dependence of the Higgs spectra on these
parameters. In Figure~\ref{fig:Higgsmasses}, we show the functional
dependence of the two lightest scalar and the lightest pseudoscalar
masses on \(\mue\) for several values of \(\mu\). The heavy states are
mainly dominated by the input \(m_{H^\pm} = 800\,\GeV\).

\begin{figure}
\begin{center}
\includegraphics[width=0.7\textwidth]{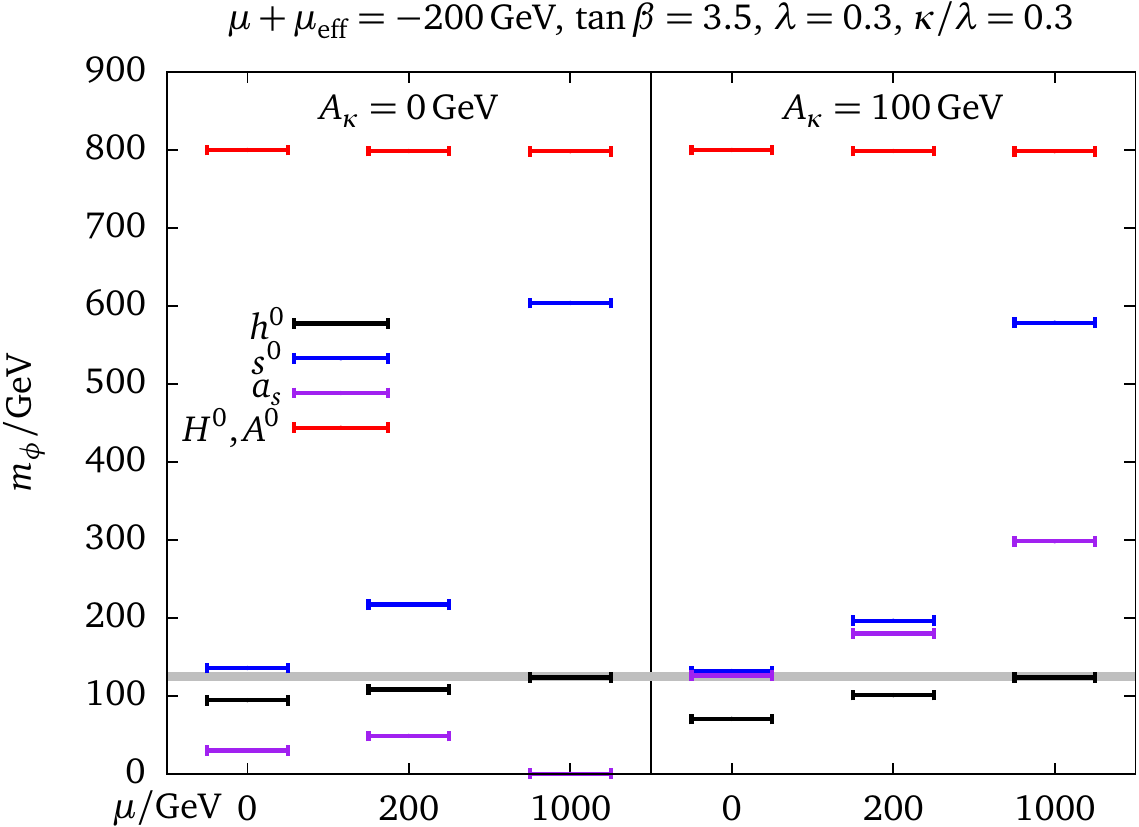}
\end{center}
\caption{The Higgs spectra change with different choice of \(\mu\) from
  \(0\), \(200\) to \(1000\,\GeV\), where the black line corresponds to
  the SM-like state. The grey band around \(125\,\GeV\) shows the
  experimentally favoured region with an error of \(\pm 3\,\GeV\). The
  effective higgsino parameter was fixed to a value of \(\mu + \mue =
  -200 \,\GeV\). On the right side, with respect to the left side, the
  \(A_\kappa\) contribution is risen from \(0\) to \(100\,\GeV\) which
  lifts the pseudoscalar singlet mass of \(a_s\) up and turns the
  tachyonic point at \(A_\kappa = 0\,\GeV\) and \(\mu = 1000\,\GeV\)
  non-tachyonic with a rather large \(a_s\) mass.}
\label{fig:Higgsspec}
\end{figure}

A precise knowledge of the Higgs sector in the NMSSM hence allows to
distinguish between the pure \(\mathds{Z}_3\)-symmetric NMSSM and the
inflation-inspired iNMSSM with the additional \(\mathds{Z}_3\)-breaking
\(\mu\)-term. So far, we have not considered the additional soft SUSY
breaking bilinear and kept it zero. In combination with a measurement of
the neutralino sector, which in contrast rather mimics the NMSSM, there
is a clear smoking gun of inflation that can be detected at an
electroweak precision machine like a future Linear Collider. The Higgs
spectrum varies severely with varying \(\mu\) as we show in
Figure~\ref{fig:Higgsspec}. Here, we compare the light pseudoscalar case
with vanishing \(A_\kappa\) with the heavier scenario where \(A_\kappa =
100\,\GeV\) for illustrative reasons. While the light pseudoscalar mass
is lifted up mainly by the amount of \(A_\kappa\), the tachyonic state
for \(\mu = 1000\,\GeV\) gets non-tachyonic and the scalar spectrum only
changes marginally. The heavy states are merely fixed by the input value of
the charged Higgs mass \(m_{H^\pm} = 800\,\GeV\).

\begin{figure}
\begin{center}
\includegraphics[width=0.7\textwidth]{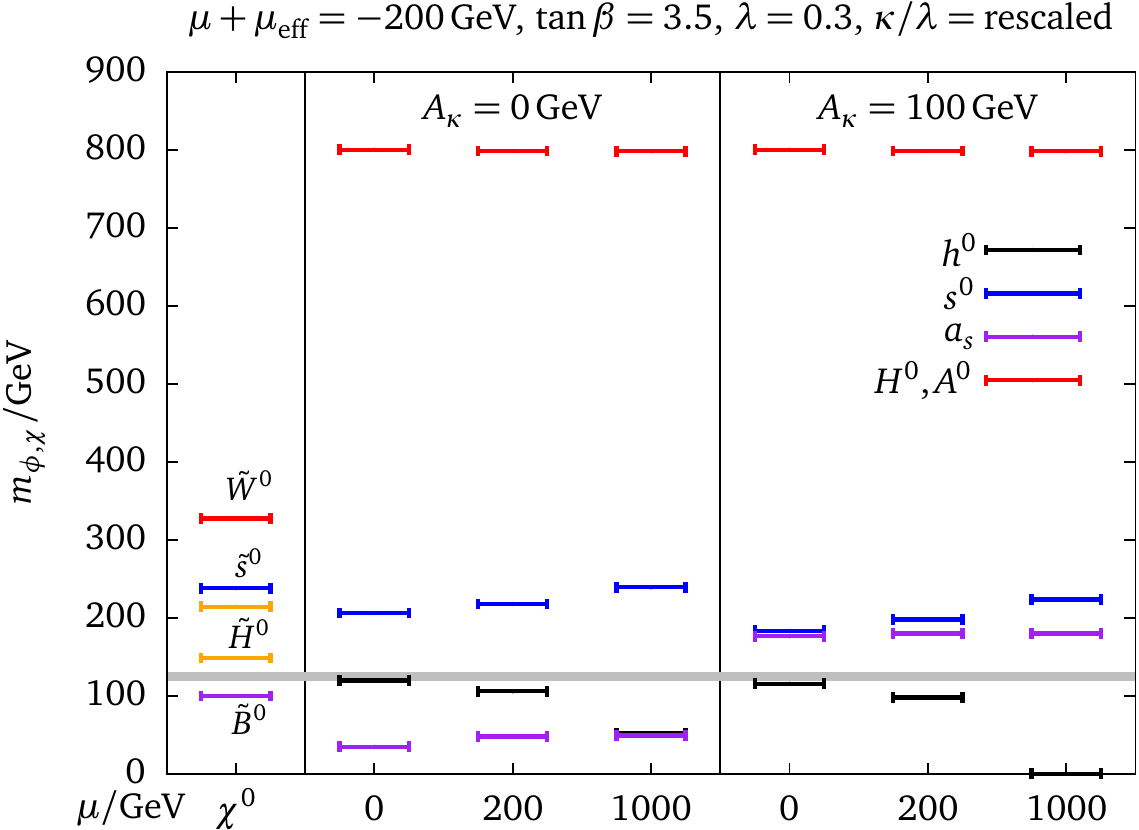}
\end{center}
\caption{While the neutralino spectrum stays invariant under change of
  \(\mu\), where the sum \(\mu + \mue = -200 \, \GeV\) is fixed and the
  ratio \(\kappa / \lambda\) rescaled in such a way to keep the singlino
  mass fixed as well, the scalar and pseudoscalar Higgs spectrum still
  change with varying \(\mu\).}
\label{fig:Higgsresc}
\end{figure}

The electroweakino sector is defined and briefly described in
Appendix~\ref{app:defin}, where it can be seen from the neutralino mass
matrix that the singlino mass is governed by \(\tfrac{\kappa}{\lambda}
\mue\), where the higgsino mass is determined by \(\mu + \mue\). Thus, a
small higgsino mass, and therefore especially also a small charged
higgsino mass, which is preferably detectable at a Linear Collider, is
somewhat naturally selected in the iNMSSM where \(\mu\) and \(\mue\)
have to have opposite signs and rather the same magnitude. Such a
cancellation, however, if \(\mu\) is significantly large, tends to
produce a heavy singlino in the iNMSSM in contrast to the NMSSM. This
effect can be removed by adjusting the ratio \(\kappa / \lambda\) in
such a way that both singlino and higgsino masses scale the same with
\(\mu\). By this redefinition, however, if \(\lambda\) is kept fixed,
the value of \(\kappa\) changes dramatically. While the electroweakino
sector may look the same as in the NMSSM even in the presence of a large
\(\mu\)-term, the (pseudo)scalar sector still has a strong dependence on
the additional \(\mu\)-term which is shown in
Figure~\ref{fig:Higgsresc}.

All the considerations and predictions presented above still depend on
other, previously suppressed parameters. For the illustrative purpose,
these additional parameters have been fixed to some values, variation of
them also changes the structure of the plots shown in this
talk. Especially the top/stop sector enters the determination of the
SM-like Higgs mass, where we show mass contours in an interesting slice
of parameter space in the following. The Higgs mass predictions contain
the full iNMSSM one-loop and leading two-loop contributions, where the
stop contribution in all cases was fixed to be sizeable and beyond the
direct limits on stop searches in such a way that \(m_{\tilde t} =
2\,\TeV\) and the mixing was chosen to be \(A_t = 2 m_{\tilde t}\). This
particular choice, of course, can and has to be adjusted in a precision
analysis. Moreover, the influence of the other input parameters as
\(\tan\beta\), \(\lambda\), \(\kappa\), and to some extend \(A_\kappa\),
still has to be tackled down in order to clearly determine the precision
needed to distinguish two different scenarios of the NMSSM and the
iNMSSM generating similar spectra. In addition, the electroweak
phenomenology also involves production and decay rates of the Higgs
states and thus one has an additional handle to distinguish the two
models. In any case, a precise measurement of the electroweak sector at
a future collider will give clear insights whether there is a smoking
gun of inflation at the Linear Collider or not. This will be discussed
in a forthcoming publication~\cite{Hollik:2018tbd}.

We have discussed above that an interesting slice of parameter space is
defined by the sum of the two \(\mu\)-terms, \(\mu + \mue\), and the
ratio \(\kappa / \lambda\). The couplings \(\lambda\) and \(\kappa\) are
known to run into a Landau pole below the GUT scale in the NMSSM, and
the same is true for the iNMSSM since the additional \(\mu\)-term does
not change the running. This non-perturbativity can be avoided, if
\(\lambda\) and \(\kappa\) are taken to be constrained by \(\lambda^2 +
\kappa^2 \lesssim 0.5\), which will be always the case in the region
plots shown in the following. The phenomenologically interesting regions
are those where the SM-like Higgs state can accommodate for the observed
\(125\,\GeV\), the branching ratios are in those cases SM-like as
well. There is an experimental exclusion from direct higgsino searches
which constrains the chargino mass to be \(\lesssim 94\,\GeV\); this
bound approximately transfers to \(\mu + \mue\).

\paragraph{Parameter scans and results}
The parameter space of the iNMSSM gets enlarged by two dimensions
(\(\mu\) and \(B_\mu\)) with respect to the NMSSM. Additionally, the new
parameters may invalidate certain allowed regions of the NMSSM that
become \eg tachyonic once a sufficiently large \(\mu\) parameter is
turned on. On the other hand, as we have seen, there might be
cancellations between \(\mu\) and \(\mue\). The surviving parameter
space appears to be rather constrained, which allows to have some clear
predictions, especially on the hierarchy of Higgs boson
masses. Unfortunately, as there are many parameters available,
modification of one of these where the others remain fixed, relax the
constraints and thus diminish the predictability. This, however, comes
along with a different phenomenology and thus clearly distinguish
different scenarios.

\begin{figure}
\begin{center}
\includegraphics[width=0.5\textwidth]{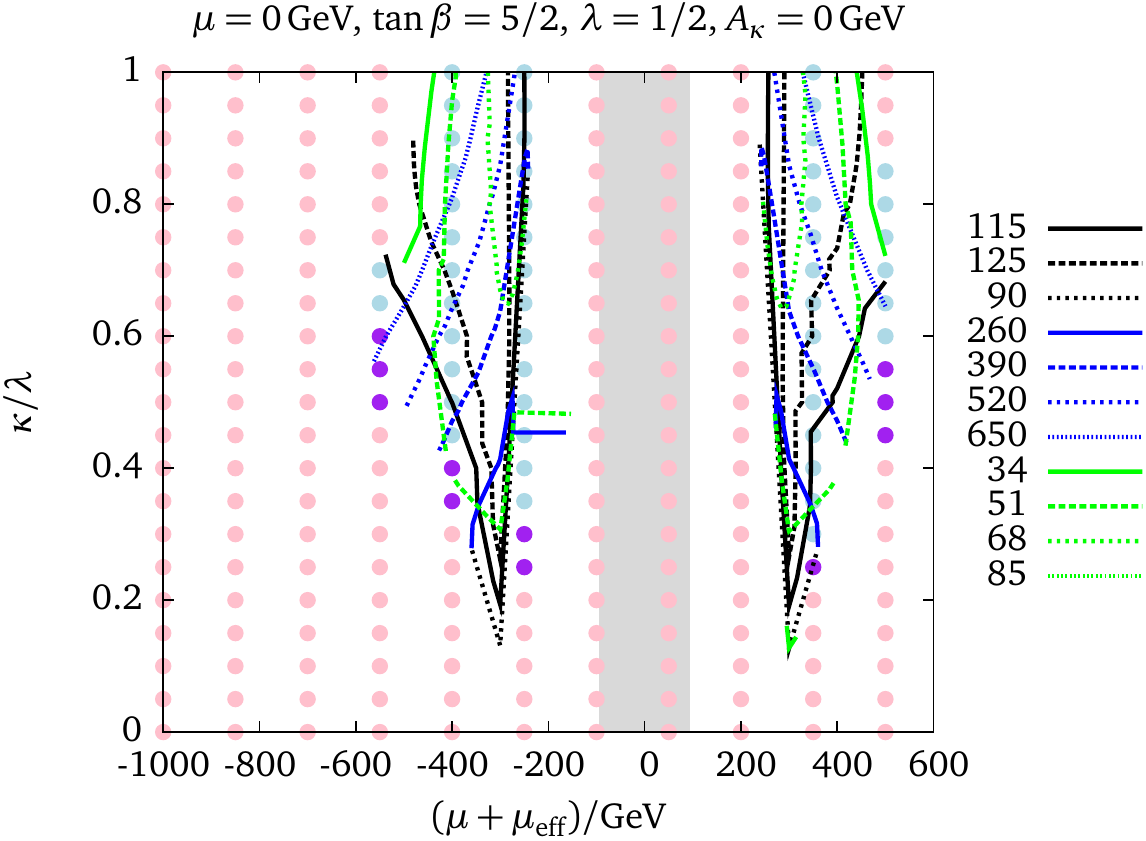}
\end{center}
\begin{minipage}{0.5\textwidth}
\includegraphics[width=\textwidth]{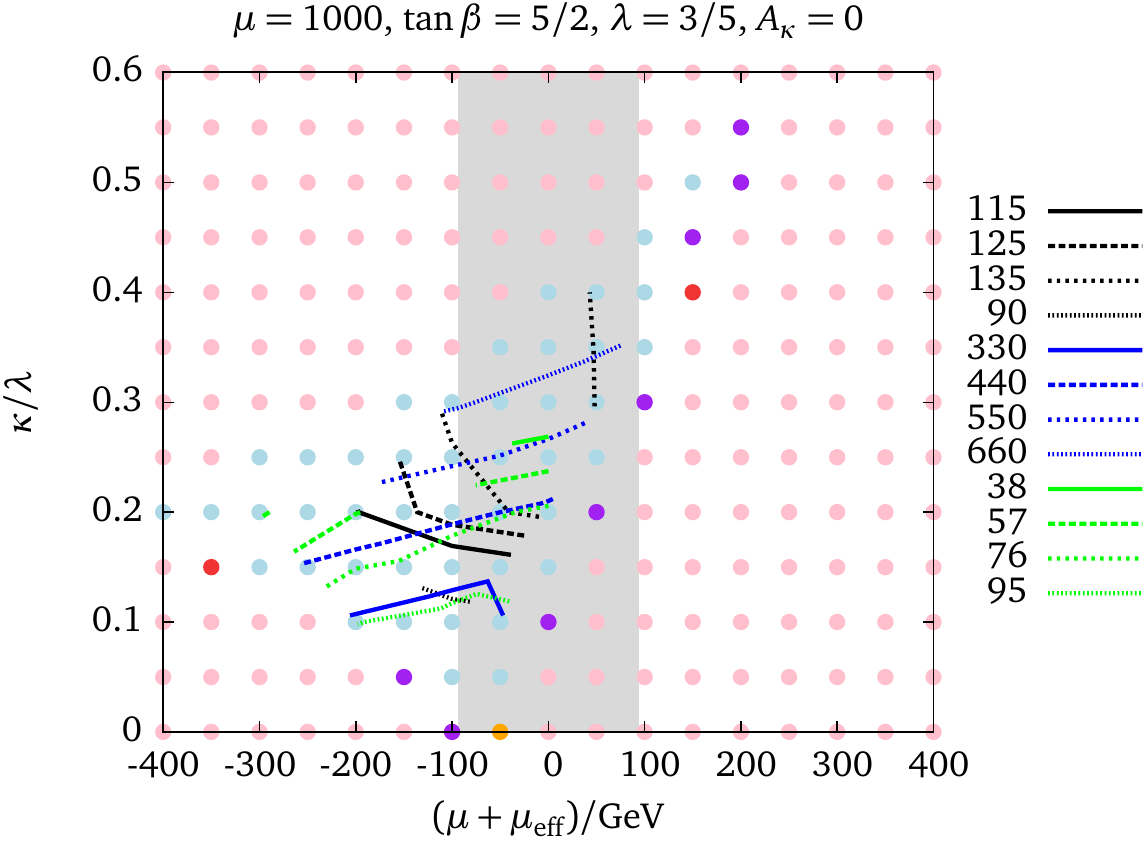}
\end{minipage}%
\begin{minipage}{0.5\textwidth}
\includegraphics[width=\textwidth]{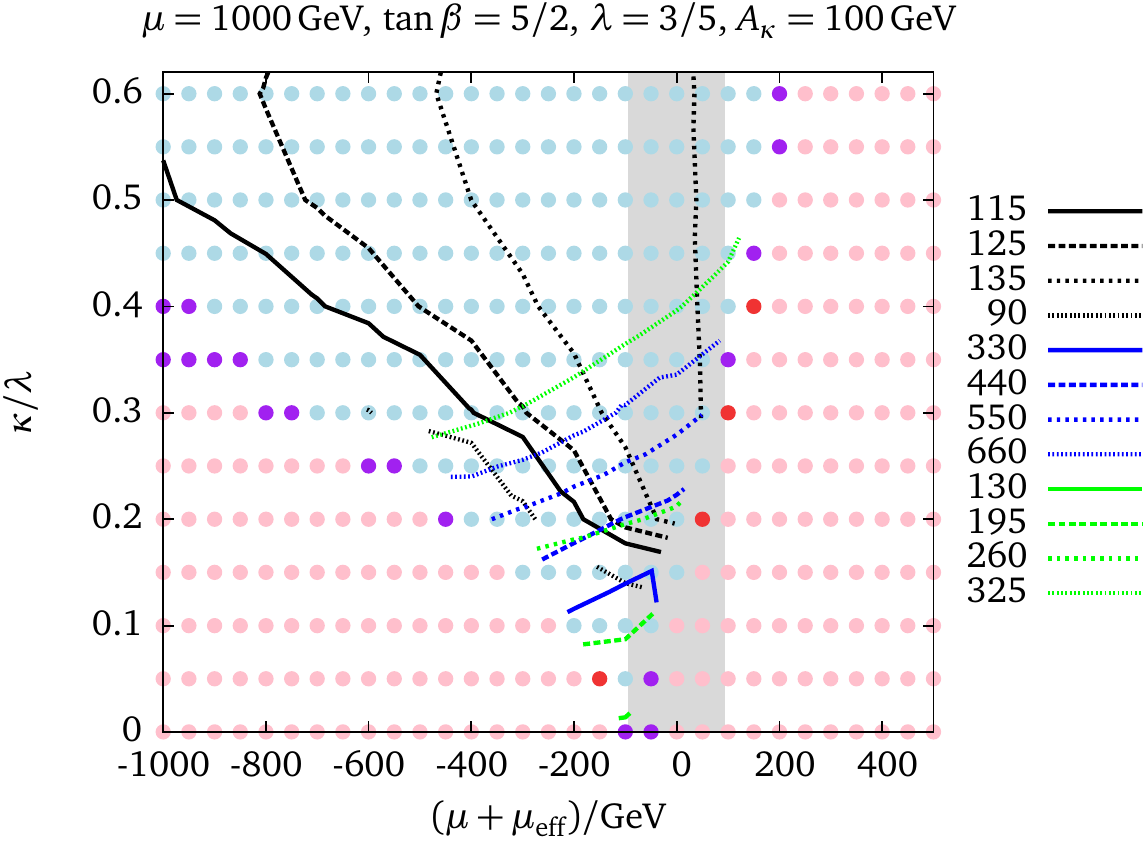}
\end{minipage}\\
\begin{minipage}{0.5\textwidth}
\includegraphics[width=\textwidth]{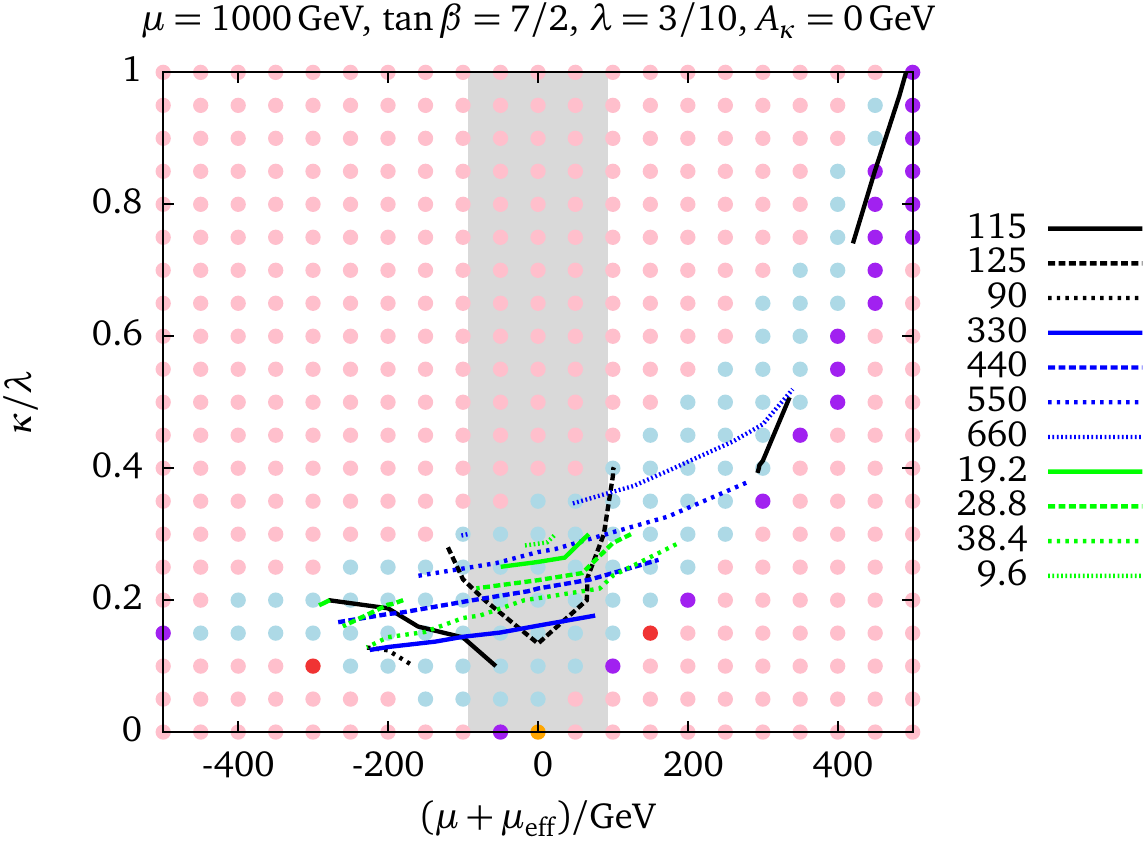}
\end{minipage}%
\begin{minipage}{0.5\textwidth}
\includegraphics[width=\textwidth]{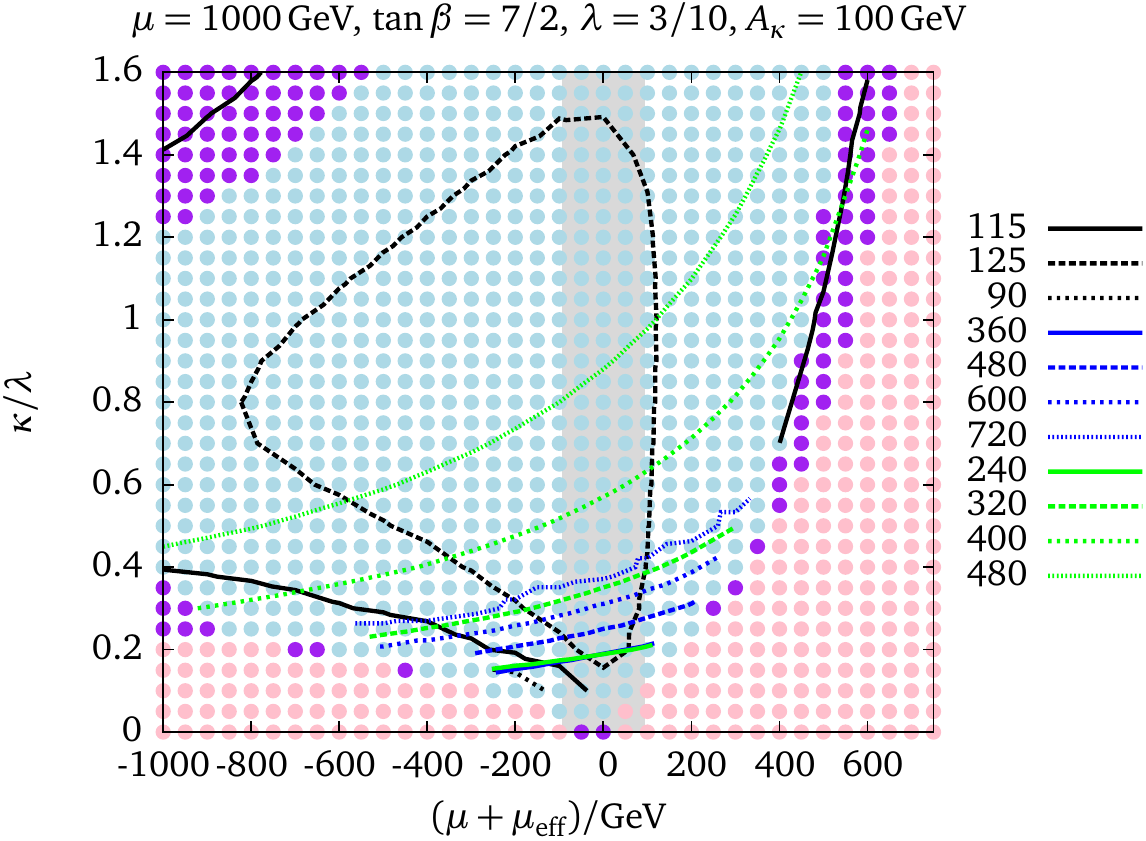}
\end{minipage}
\caption{This selection shows some typical features of the iNMSSM and
  NMSSM. The upper plot has the situation with a vanishing \(\mu\) and
  \(A_\kappa\) which is symmetric around \(\mue = 0\,\GeV\). The lower
  plots have a larger \(\mu = 1\,\TeV\) and show the effect of
  \(A_\kappa\) on the allowed space and the Higgs masses. The contour
  lines give lines of equal masses for the SM-like Higgs (black), the
  singlet-like scalar (blue) and singlet-like pseudoscalar (green) which
  can be light for small \(A_\kappa\). The colour code for the points
  describes the stability of the electroweak ground state: light blue
  points have a global electroweak vacuum, purple ones a long-lived and red
  points a short-lived desired vacuum. Tachyonic points are shown in
  pink, where the orange points do not fulfil the NMSSM constraint
  \(A_\kappa^2 > 9 m_S^2\) for a non-vanishing singlet \vev.}
\label{fig:sampleplots}
\end{figure}

One crucial parameter, as already discussed above, is given by
\(A_\kappa\) which controls the mass of the singlet pseudoscalar
state. Low values of \(A_\kappa\) produce a rather light state, heavier
masses can be generated by lifting \(A_\kappa\) up and simultaneously
removing tachyons from the spectrum. There appears to be a larger
fraction in the parameter space allowed, if one takes a look at the
\(\kappa/\lambda\) vs.\ \(\mu+\mue\) slice. This is shown in the samples
of Figure~\ref{fig:sampleplots}. Comparing the cases with \(A_\kappa =
0\) and \(100\,\GeV\), the effect of opening up excluded tachyonic
parameter space can be clearly seen for the cost of a heavier singlet
pseudoscalar (green contours). The other parameters only have a minor
effect, so \(\tan\beta\) is enhanced from \(2.5\) to \(3.5\) from the
second to the third row of Figure~\ref{fig:sampleplots} and
simultaneously \(\lambda\) reduced from \(0.6\) to \(0.3\), where \(\mu
= 1\,\TeV\) in all cases. The single plot on top of
Figure~\ref{fig:sampleplots} illustrates the ``NMSSM-limit'' with
vanishing \(\mu\). Here, apparently only a very constrained region is
allowed (note that \(A_\kappa = 0\,\GeV\)) and the singlet-like
pseudoscalar can be rather light. The grey bands show a rough
experimental exclusion on the higgsino mass given by the LEP-limit on
the chargino mass \(m_{\chi^\pm_1} > 94\,\GeV\)
\cite{Patrignani:2016xqp}. The chargino mass in the iNMSSM is mainly
given by \(\mu+\mue\), see Appendix~\ref{app:defin}, up to small
modifications from the mixing.

Figure~\ref{fig:sampleplots} also reveals information about the vacuum
structure of the scanned points: light blue points denote an absolutely
stable electroweak vacuum, where tachyonic states (at the tree-level!)
appear in the pink points. Interestingly, the allowed regions can also
easily accommodate for a \(125\,\GeV\) SM-like Higgs state, where we
added a uniform stop contribution as discussed above. Unstable or
metastable desired vacua are coded in purple (long-lived) and red
(short-lived). We briefly describe in Appendix~\ref{app:vactun} how we
estimate the life-time. In the NMSSM there exists a bound on
\(A_\kappa\),
\begin{equation} \label{eq:Akapbound}
A_\kappa^2 > 9 m_S^2,
\end{equation}
relating the trilinear soft SUSY breaking singlet coupling with the soft
SUSY breaking singlet mass. This constraint is needed to generate a
sufficiently large singlet \vev{} and therefore higgsino mass. In the
presence of the \(\mathds{Z}_3\)-breaking \(\mu\)-term, this unequation
does not have to be necessarily fulfiled. Eq.~\eqref{eq:Akapbound} can
be easily derived from the singlet-only potential with the requirement
that the minimum \(\langle S \rangle \neq 0\) is the true vacuum and
thus a non-vanishing singlet \vev{} is generated. This is needed in the
\(\mathds{Z}_3\)-invariant NMSSM to produce the correct electroweak
phenomenology. In the iNMSSM, however, the \(\mathds{Z}_3\)-breaking
MSSM-like \(\mu\)-term is generated by the non-minimal coupling to
supergravity and related to the scale of SUSY breaking and the gravitino
mass. If both \(\mu\) and \(\mue\) are present, there can be
cancellations since they have to have different signs and hence a small
higgsino mass still can be valid even if both \(\mu\) parameters are in
the \(\TeV\) range.

\paragraph{Constraints on \(B_\mu\)}
The iNMSSM has in addition to the superpotential parameter one more soft
SUSY breaking term, the bilinear \(B_\mu\)-term, which has been ignored
to far in the discussion above. It turns out that it cannot be
arbitrarily large anyway and thus there are good reasons to keep it
small. If it is non-zero, the effect is merely under control as the
contribution from \(B_\mu\) grows linearly with \(\mu\) (note that it
appears as \(\mu B_\mu\) in the soft breaking potential). Together with
\(A_\lambda\) it influences the charged Higgs mass and therefore, in our
approach where we treat \(A_\lambda\) for \(m_{H^\pm}\) as input, it
enters the determination of \(A_\lambda\), see Eq.~\eqref{eq:Alambda}
and Appendix~\ref{app:defin}.

\begin{figure}
\begin{minipage}{0.5\textwidth}
\includegraphics[width=\textwidth]{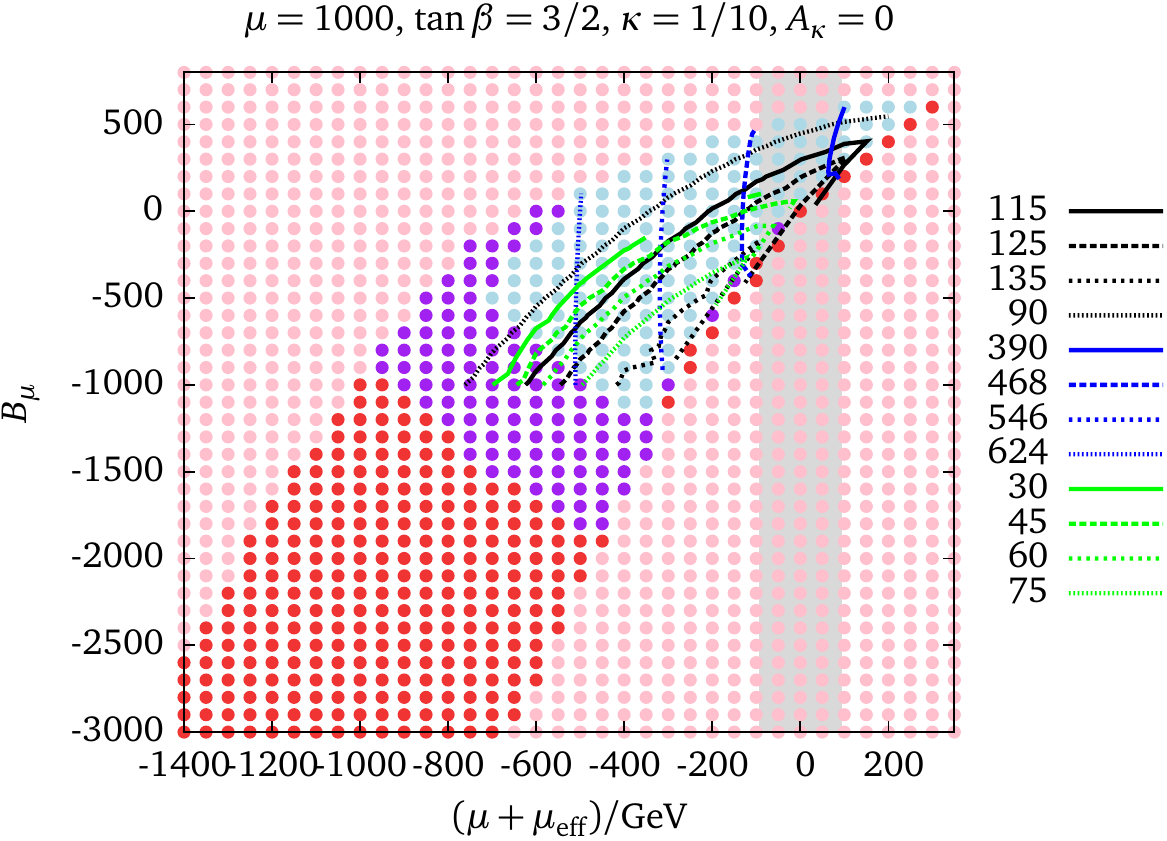}
\end{minipage}%
\begin{minipage}{0.5\textwidth}
\includegraphics[width=\textwidth]{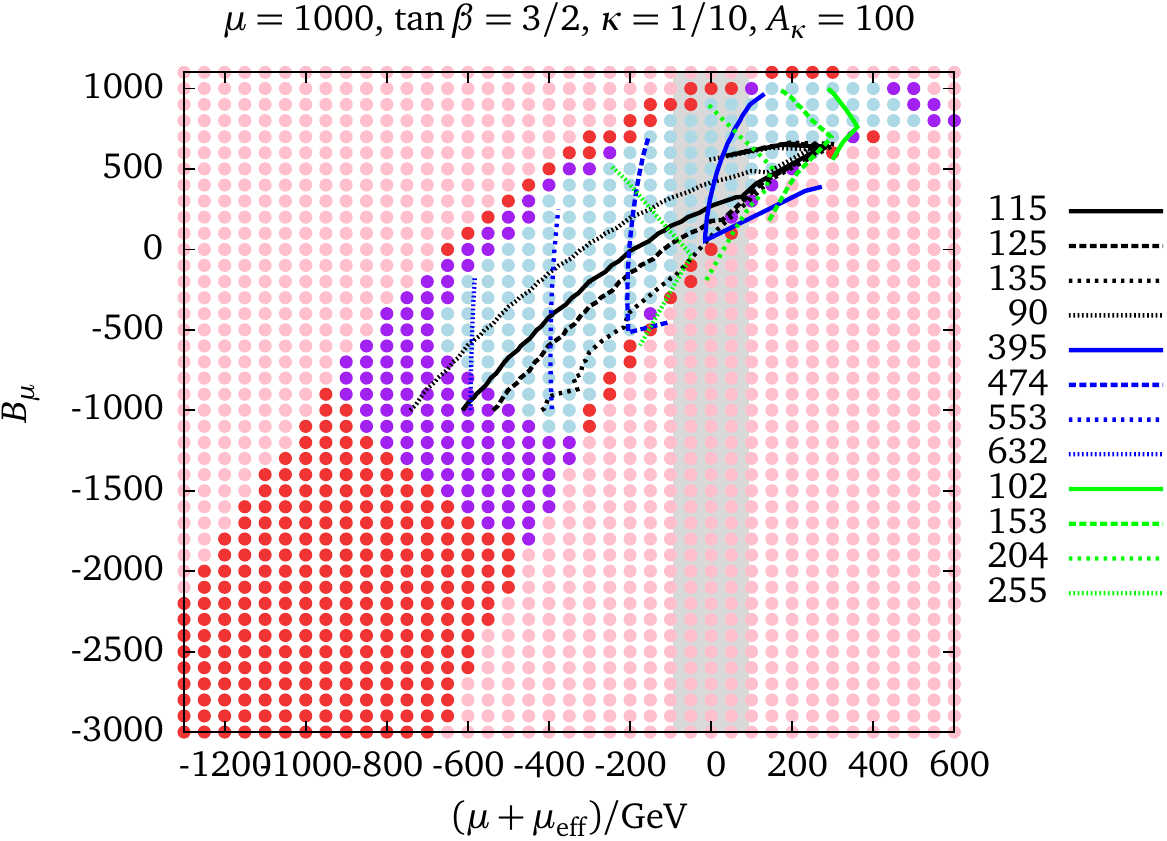}
\end{minipage}
\caption{The color code of the points and lines is as in
  Figure~\ref{fig:sampleplots}. Now we show for fixed values of
  \(\kappa = 0.1\) and \(\lambda = 0.5\) a different slice in parameter space
  for non-vanishing \(B_\mu\). In this region, the influence from
  \(A_\kappa\) is apparently very mild (left \(A_\kappa = 0\,\GeV\),
  right \(A_\kappa = 100\,\GeV\)). For too large negative values of
  \(B_\mu\), the electroweak vacuum gets severely short-lived (red
  points) with a broad band of long-lived desired vacua in between
  (purple).}
\label{fig:Bmu}
\end{figure}

Unfortunately, the role of \(B_\mu\) is less clear than compared to the
MSSM where it can easily be replaced by the pseudoscalar mass
\(m_A\). However, its impact on the Higgs boson masses is very
well-defined as it always enters in sum with \(\mue
\tfrac{\kappa}{\lambda}\) and \(\mue A_\lambda\). Thus, it might be
absorbed in \(A_\lambda\), which nevertheless appears also at different
places. Treating the charged Higgs mass \(m_{H^\pm}\) as input and
solving for \(A_\lambda\), \(\mu B_\mu\) enters the determination of
\(A_\lambda\). For too large values of \(B_\mu\), tachyonic states are
generated again. There is, however, a valley that allows for
non-tachyonic states even for large but negative \(B_\mu\) values as can
be seen from Figure~\ref{fig:Bmu}. It has nevertheless the power to
destabilize the desired electroweak ground state of the theory as such
large values of \(B_\mu\) induce a global minimum different from the
standard vacuum. The desired but local vacuum now appears to be rather
short-lived with respect to the life-time of the universe, which is
depicted by the red points of Figure~\ref{fig:Bmu}. In the boundary
region, the electroweak vacuum is sufficiently long-lived (purple
points). The scalar singlet mass (blue lines) appears to be rather
independent of \(B_\mu\), where the pseudoscalar singlet (green) shows a
striking behaviour depending on \(A_\kappa\). The mass of the SM-like
Higgs boson is very much aligned with the allowed valley and explains
very well the tachyonic boundary (together with the pseudoscalar singlet
as can be already seen from Figure~\ref{fig:Higgsmasses}, where \(B_\mu
= 0\,\GeV\) but the two light states running tachyonic at different
places).

\section{Conclusions}\label{sec:concl}
We have presented the electroweak phenomenology of an inflation-inspired
NMSSM as first discussed in Refs.~\cite{Einhorn:2009bh, Ferrara:2010yw,
  Ferrara:2010in}. We briefly summarized the idea of Higgs inflation in
the superconformal sector and showed how the non-minimal coupling of the
Higgs sector to supergravity shows up in the effective low-energy
superpotential. The remaining model can be described by the NMSSM
augmented with an MSSM-like \(\mu\)-term \(\mu\; H_u \cdot H_d\), which
breaks the accidental \(\mathds{Z}_3\) invariance of the
NMSSM. Additionally, a soft SUSY breaking term \(\mu B_\mu\) is
generated and has to be taken into account. The rules of supergravity
dictate \(\mu\) and together with the effective \(\mu\)-term of the
NMSSM arising from the singlet \vev{}, both sum up to an effective
higgsino mass \(\mu + \mue\). This combination plays an important
role for the phenomenology, especially since the signs of both
contributions appear to be anticorrelated and thus a natural
cancellation among those fundamentally different contributions to the
higgsino mass appears. Thus, scenarios with light higgsinos but heavy
singlinos exist and a precise knowledge of the Higgs and electroweakino
sector as it might be achieved at a future Linear Collider helps to
clearly distinguish this model from the ordinary NMSSM. A smoking gun of
the inflationary remnant exists as a footprint in the electroweak
spectrum.

We have extensively discussed the influence of the model parameters on
the Higgs masses and how tachyonic states are generated at the
tree-level. Tachyonic masses invalidate the expansion point in such a
way that the electroweak point appears to be a local maximum instead of
a minimum and thus the tachyonic direction points towards the global
minimum. In addition, the iNMSSM as well as the NMSSM may reveal several
vacua out of which the desired vacuum appears to be a false vacuum. A
numerical analysis minimising the scalar potential finds the global
minimum of the theory, which in some cases is the electroweak vacuum in
others not. If the desired vacuum is a local minimum, vacuum decay rates
have been estimated to compare the life-time of the false vacuum with
the life-time of the universe. Only in the case of large and negative
\(B_\mu\) values, reasonable amounts of short-lived vacua have been
found.

Higgs inflation embedded into a superconformal framework appears to be
distinguishable at low energies from the common SUSY models beyond the
SM. The iNMSSM needs an additional singlet as the NMSSM; the spectrum,
however, appears to be different and cannot be matched to the parameters
of the NMSSM. The model is also different from the MSSM, in which Higgs
inflation cannot be accommodated.

The results presented in this talk are going to be discussed in more
detail in a forthcoming publication~\cite{Hollik:2018tbd}.

\section*{Acknowledgements}
The speaker thanks his collaborators S.~Liebler, G.~Moortgart-Pick,
S.~Passehr and G.~Weiglein for their contribution. This work is
supported by the Deutsche Forschungsgemeinschaft through a lump sum fund
of the SFB 676 ``Particles, Strings and the Early Universe''.

\appendix

\section{Higgs boson and neutralino/chargino mass matrices}
\label{app:defin}
We define the Higgs mass matrices via the second derivatives of the
potential, where we distinguish between scalar and pseudoscalar neutral
states by the decomposition
\begin{equation}
\begin{aligned}
H_u &= \begin{pmatrix} H_u^+ \\ H_u^0 \end{pmatrix} =
\begin{pmatrix} \phi^+_u \\ v_u + \tfrac{1}{\sqrt{2}} \left( \sigma_u + \im
    \phi_d \right) \end{pmatrix}, \qquad
H_u = \begin{pmatrix} H_d^0 \\ H_d^- \end{pmatrix} =
\begin{pmatrix} v_d + \tfrac{1}{\sqrt{2}} \left( \sigma_d + \im \phi_d
  \right) \\ \phi_d^- \end{pmatrix}, \\
S &= v_s + \tfrac{1}{\sqrt{2}} \left(\sigma_s + \im \phi_s \right).
\end{aligned}
\end{equation}
The mass matrices for the scalar and pseudoscalar states
\(\mathcal{M}^2_S\) and \(\mathcal{M}^2_P\), respectively, are then
given by the expressions
\begin{subequations}
\begin{align}
\mathcal{M}^2_S &= \begin{pmatrix}
M_Z^2 c_\beta^2 + a_1 t_\beta & \left( 2v^2\lambda^2 - M_Z^2 \right)
c_\beta s_\beta - a_1 & a_2 c_\beta - a_3 s_\beta \\
* & M_Z^2 s_\beta^2 + a_1 / t_\beta & a_2 s_\beta - a_3 c_\beta \\
* & * & a_4 + a_5
\end{pmatrix}, \\
\mathcal{M}^2_P &= \begin{pmatrix}
a_1 t_\beta & a_1 & - a_6 s_\beta \\
* & a_1 / t_\beta & - a_6 c_\beta \\
* & * & a_4 - 3 a_5 - 2 a_7
\end{pmatrix},
\end{align}
\end{subequations}
with \(s_\beta = \sin\beta\), \(c_\beta = \cos\beta\), \(t_\beta =
\tan\beta\) and where the abbreviations \(a_i\) are
\begin{subequations}
\begin{align}
a_1 &= B_\mu \mu + \mue \left( \frac{\kappa}{\lambda} \mue + A_\lambda
\right), \\
a_2 &= 2 v \lambda \left( \mu + \mue \right), \\
a_3 &= v \lambda \left( 2 \frac{\kappa}{\lambda} \mue + A_\lambda
\right), \\
a_4 &= \frac{1}{\mue} \left[ v^2 \lambda^2 c_\beta s_\beta \left(
    \frac{\kappa}{\lambda} \mue + A_\lambda \right) - v^2 \lambda^2 \mu
\right], \\
a_5 &= 4 \left( \frac{\kappa}{\lambda} \right)^2 \mue^2 +
\frac{\kappa}{\lambda} \left[ \mue \, A_\kappa - v^2 \lambda^2 c_\beta
  s_\beta \right], \\
a_6 &= v \lambda \left( 2 \frac{\kappa}{\lambda} \mue - A_\lambda
\right), \\
a_7 &= -6 \left( \frac{\kappa}{\lambda} \right)^2 \mue^2 \;.
\end{align}
\end{subequations}
Note, that the pseudoscalar mass matrix comprises one massless state,
the Goldstone mode. The charged Higgs mass matrix is given by
\begin{equation}
\mathcal{M}^2_C = \left[ \left(M_W^2 - v^2 \lambda^2 \right) c_\beta
  s_\beta + a_1 \right]
\begin{pmatrix}
  t_\beta & 1 \\ 1 & 1 / t_\beta
\end{pmatrix},
\end{equation}
with the \(W\) boson mass \(M_W^2 = \tfrac{1}{2} g_2^2 v^2\) and the
eigenvalue given by
\begin{equation}
m_{H^\pm} = M_W^2 - v^2 \lambda^2 + \frac{a_1}{c_\beta s_\beta},
\end{equation}
which can be used to eliminate \(A_\lambda\) as a free parameter for the
sake of the charged Higgs boson mass \(m_{H^\pm}\) as input value (for
the numerical analyses presented in this talk, we used allover the
scenarios \(m_{H^\pm} = 800 \, \GeV\)), such that
\begin{equation} \label{eq:Alambda}
A_\lambda = \frac{c_\beta s_\beta}{\mue} \left( m_{H^\pm}^2 - M_W^2 +
  v^2 \lambda^2 \right) - \frac{B_\mu \mu}{\mue} - \mue
\frac{\kappa}{\lambda}.
\end{equation}

The mass matrices of charginos and neutralinos resemble very much the
ordinary NMSSM, where the effective higgsino parameter is replaced by
\(\mu + \mue\). However, the singlino mass is only governed by \(\mue\),
since the additional \(\mu\) term couples the doublet
superfields. Therefore, the neutralino mass matrix is given by
\begin{equation} \label{eq:neutralino}
\mathcal{M}_{\chi^0} = \begin{pmatrix}
M_1 & 0 & - M_Z s_w c_\beta & M_Z s_w s_\beta & 0 \\
* & M_2 & M_Z c_w c_\beta & - M_Z c_w s_\beta & 0 \\
* & * & 0 & - (\mu + \mue) & - \lambda v s_\beta \\
* & * & * & 0 & - \lambda v c_\beta \\
* & * & * & * & 2 \tfrac{\kappa}{\lambda} \mue
\end{pmatrix},
\end{equation}
where \(M_1\) and \(M_2\) are the gaugino masses for the \(\U(1)_Y\) and
\(SU(2)_L\) gauginos, respectively, and the weak mixing angle
\(\theta_w\) enters via \(\tan\theta_w = s_w / c_w = g_1 /
g_2\). Apparently, the neutralino spectrum in the iNMSSM can be rescaled
via the ratio \(\tfrac{\kappa}{\lambda}\) in such a way to match the
NMSSM neutralino spectrum for a given higgsino mass \(\mu + \mue\).

The chargino mass matrix is given by
\begin{equation}
\mathcal{M}_{\chi^\pm} = \begin{pmatrix}
M_2 & \sqrt{2} M_W s_\beta \\
\sqrt{2} M_W c_\beta & \mu + \mue
\end{pmatrix}.
\end{equation}

\section{Vacuum tunneling} \label{app:vactun}
We briefly describe our estimate on the tunneling rates in the case
where the desired vacuum appears to be a false vacuum. The electroweak
input parameters determine the position and depth of the local minimum,
where the global minimum and true vacuum is found by numerical
minimisation of the tree-level potential. In general, the true vacuum
has \vev{}s \(\langle H^0_u \rangle \neq v_u\), \(\langle H^0_d \rangle
\neq v_d\) and \(\langle S \rangle \neq v_s\). We approximate the
potential barrier between the two minima by a one-dimensional quartic
potential potential
\begin{equation} \label{eq:onedimpot}
V(\phi) = g\; \phi^4 - a\; \phi^3 + b\; \phi^2 + c\; \phi + d,
\end{equation}
which allows for an exact solution of the bounce
action~\cite{Adams:1993zs}. This is given by
\begin{equation}
B = \frac{\pi^4}{3 g} (2 - \delta)^{-3} \left[ \alpha_1 \delta +
  \alpha_2 \delta^2 + \alpha_3 \delta^3 \right],
\end{equation}
with \(\delta = 8 g^2 b / a^2\) and \(\alpha_{1,2,3}\) numerical
coefficients. By comparison of the decay rate of the false vacuum per
unit volume~\cite{Coleman:1977py}
\begin{equation}
\Gamma / V = A e^{-B / \hbar} \left[ 1 + \mathcal{O}(\hbar) \right],
\end{equation}
one estimates bounce actions \(B \gtrsim 400\) to be sufficiently
long-lived. The prefactor \(A\) is difficult to calculate and usually
approximated by the height of the potential or the electroweak scale,
\(A \sim (100\,\GeV)^4\), where the error enters only logarithmically the
decay time.

The interpolation between the two minima is done by a straight line,
where the electroweak point is shifted to the origin. Therefore, in the
expression of the neutral Higgs potential, we have the replacement
\begin{equation}
V(\phi) = V\left( H_u^0 = v_u + (V_u - v_u) \frac{\phi}{\sqrt{2}}, H_d^0
  = v_d + (V_d - v_d) \frac{\phi}{\sqrt{2}}, S = v_s + (V_s - v_s)
  \frac{\phi}{\sqrt{2}} \right),
\end{equation}
with the ``true'' \vev{}s \(V_u\), \(V_d\) and \(V_s\). The factor
\(1/\sqrt{2}\) is employed to keep the \(\phi\)-field canonically
normalised. This way, the field \(\phi\) interpolates between the
desired, false vacuum (\(\phi = 0\)) and the true vacuum (\(\phi =
\sqrt{2}\)). It is, however, more convenient to keep \(\phi\) dimensionful and
thus the coefficients of the potential in Eq.~\eqref{eq:onedimpot} of
the same order of magnitude as the original coefficients. Therefore, we
use a normalised field in the one-field potential, \(V(\overline\phi)\), with
\begin{equation}
\overline \phi = \frac{\phi}{\sqrt{(V_u-v_u)^2 / 2 + (V_d-v_d)^2 / 2 + (V_s -
    v_s)^2 / 2}}.
\end{equation}

\bibliographystyle{utcaps}
\bibliography{hinfl}

\end{document}